\documentclass[singlecolumn]{elsarticle}

\pdfoutput=1

\usepackage{graphicx}
\usepackage{amsmath,amssymb}

\journal{Physics Letters B}
\date{hello}

\newcommand{\nn}{\nonumber} 
\newcommand{\bea}{\begin{eqnarray}}
\newcommand{\eea}{\end{eqnarray}}
\newcommand{\beq}{\begin{equation}}
\newcommand{\eeq}{\end{equation}}

\newcommand{\phie}{\sigma}
\newcommand{\phid}{\psi}
\newcommand{\kin}{k}
\newcommand{\mpl}{m_{\mbox {\tiny Pl}}}

\newcommand{\sfun}{s}

\begin{document}

\begin{frontmatter}



\title{Running Inflation in the Standard Model}

\author{Andrea De Simone\footnote{Electronic address: {\tt andreads} at {\tt mit.edu}},\,
Mark P.  Hertzberg\footnote{Electronic address: {\tt mphertz} at {\tt mit.edu}},\,
and Frank Wilczek\footnote{Electronic address: {\tt wilczek} at {\tt mit.edu}}
\\
~\\
Center for Theoretical Physics and Department of Physics,\\ 
Massachusetts Institute of Technology, Cambridge, MA 02139, USA}


\begin{abstract}
An interacting scalar field with largish coupling to curvature can support a distinctive inflationary universe scenario.   Previously this has been discussed for the Standard Model Higgs field, treated classically or in a leading log approximation.  Here we investigate the quantum theory using renormalization group methods.  In this model the running of both the effective Planck mass and the couplings is important.  The cosmological predictions are consistent with existing WMAP5 data, with $0.967\lesssim n_s\lesssim0.98$ (for $N_e=60$) and negligible gravity waves.   We find a relationship between the spectral index and the Higgs mass that is sharply varying for $m_h\sim 120-135$\,GeV (depending on the top mass); in the future, that relationship could be tested against data from PLANCK and LHC.   We also comment briefly on how similar dynamics might arise in more general settings, and discuss our assumptions from the effective field theory point of view.
\end{abstract}


\end{frontmatter}


\section{Introduction}
\label{sec:introduction}

The hypothesis that there was a period in the early history of the universe during which a local Lorentz invariant energy density -- i.e., an effective cosmological term -- dominated the equation of state, causing exponential expansion, explains several otherwise puzzling features of the present universe (flatness, isotropy, homogeneity) \cite{Guth,Linde,AlbrechtSteinhardt82,Linde83}.  It also suggests a mechanism whereby primordial density fluctuations arise through intrinsic fluctuations of quantum fields, leading to qualitative and semi-quantitative predictions that are consistent with recent observations.  
However the physics behind inflation remains mysterious.  What, specifically, is the source of the energy density?  Ideas ranging from fields associated with supersymmetry, string moduli, ghosts, branes, and others abound \cite{ModLinde, ModLiddle,ModDvali, ModMcAllister, ModNima, ModKachru}.    One (or more) of them might be correct, but all are highly speculative, and none is obviously compelling.

Alternatively, we can look for inflationary dynamics based on degrees of freedom already present in the Standard Model.   We can also attempt to maintain the guiding philosophy of the Standard Model, including gravity, to allow only local interactions which are gauge invariant and have mass dimension $\leq 4$.   Within this very restrictive framework, there remains the possibility to include the non-minimal gravitational coupling  $\xi H^\dagger H\mathcal{R}$.  Here $H$ is the Higgs field, $\mathcal{R}$ is the Ricci scalar, and $\xi$ is a dimensionless coupling constant, whose value is unknown and largely unconstrained by experiment.\footnote{For $\xi=-1/6$ the Higgs is conformally coupled to gravity.}
Indeed renormalization
of the divergences arising in a self-interacting scalar theory in curved spacetime requires a term of this form \cite{BirrellDavies}.  
The Higgs sector is then described, classically, by the Lagrangian
\beq
\mathcal{L}_h=
-|\partial H|^2 + \mu^2H^\dagger H-\lambda (H^\dagger H)^2+\xi H^\dagger H\,\mathcal{R}
\label{lagrangianH} \eeq
where $\lambda$ is the Higgs self coupling and 
$\mu$  is the Higgs mass parameter. 


It has been known for some time that such minimal classical Lagrangians can support inflation driven by an interesting interplay between the quartic term and the non-minimal coupling term \cite{Salopek,Fakir,Kaiser,Komatsu}.  For ease of reference, we will call this general set-up ``running inflation''; the name seems appropriate, since evolution of the effective Planck mass and the effective scalar mass is central to the dynamics.\footnote{The term ``running inflation" was used in a different context in \cite{runningpaper}.} 
This quasi-renormalizable set-up allows use of renormalization group methods, as will be illustrated here. By quasi-renormalizable, we mean that the theory is renormalizable when gravity is treated classically; in particular, we ignore quantum corrections from graviton exchange (see Appendix \ref{NewApp}). In the investigation of (non-gravitational) quantum effects, it is appropriate to focus specifically on the Standard Model, for two reasons.
 First, because (as we will see) it illustrates important qualitative issues in a very concrete, familiar setting.  Second, because it -- or something close to it -- might actually contain the degrees of freedom relevant to real-world inflation, in which case the specific predictions we derive could help describe reality.

\begin{figure}[t]
\begin{center}
\includegraphics[scale=1.0]{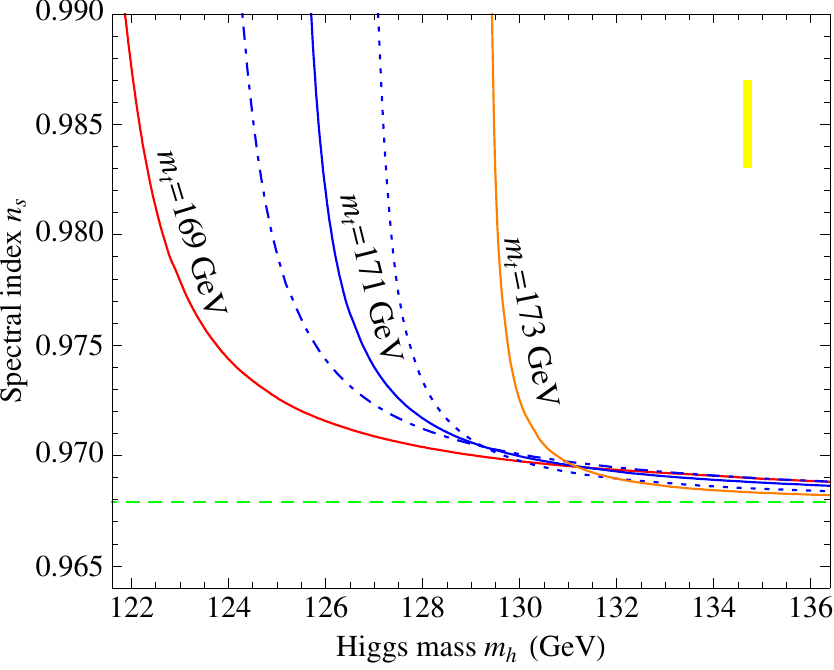}
\end{center}
\caption{The spectral index $n_s$ as a function of the Higgs mass $m_h$ for a range of light Higgs masses.
The 3 curves correspond to 3 different values of the top mass: $m_t=169$\,GeV (red curve), $m_t=171$\,GeV (blue curve), and $m_t=173$\,GeV (orange curve). The solid curves are for $\alpha_s(m_Z)=0.1176$, while for $m_t=171$\,GeV (blue curve) we have also indicated the 2-sigma spread in $\alpha_s(m_Z)=0.1176\pm0.0020$, where the dotted (dot-dashed) curve corresponds to smaller (larger) $\alpha_s$. 
The horizontal dashed green curve, with $n_s\simeq 0.968$, is the classical result. 
The yellow rectangle indicates the expected accuracy of PLANCK in measuring $n_s$ ($\Delta n_s\approx 0.004$) and the LHC in measuring $m_h$ ($\Delta m_h\approx 0.2$\,GeV).  In this plot we have set $N_e=60$.}
\label{nsplot}
\end{figure}

Recently, the idea that the Standard Model Higgs field, non-minimally coupled to gravity, 
can lead to inflation was proposed in Ref.~\cite{Bezrukov}. Those authors argued that the radiative corrections to the potential are negligible
and hence the inflationary parameters can be computed using the classical 
Lagrangian. They found that the cosmological predictions are in good agreement with cosmological data, independent of the Standard Model parameters, such as $\lambda$.
On the other hand the authors of Ref.~\cite{Barvinsky} criticized their approach,
suggesting that the quantum corrections to the potential can be very important.
They concluded that a Higgs lighter than $230$\,GeV
cannot serve as the inflaton, because the predicted spectral index is ruled out by  WMAP5 data \cite{wmap5}. 
Ref.~\cite{Barvinsky} only incorporated  quantum corrections at leading log order, extrapolated from low energies.
Here, in contrast, we will compute the full renormalization group improved effective action at 2-loops.
We conclude that running inflation based upon a Standard Model Higgs
makes predictions that are consistent with current cosmological data, and leads to firm predictions for the PLANCK satellite and the LHC.
Our main result is a correlation between the spectral index and the Higgs mass, see Fig.~\ref{nsplot}.
This correlation is absent in the classical theory. The origin of the correlation lies in the interactions of the Standard Model, which dictate the form of the effective action. 

In Section \ref{sec:nonmin} we review inflation with non-minimally coupled scalars. In Section \ref{sec:classical} we investigate the classical theory of the Higgs non-minimally coupled to gravity. In Section \ref{sec:quantum} we describe our method for obtaining the quantum corrected effective action.
We compute all the inflationary observables numerically and present results in Section \ref{sec:results}. Finally, we review our results and discuss their significance in Section \ref{sec:discussion}.

\section{Non-Minimal Inflation}
\label{sec:nonmin}

Here we briefly review the recipe to compute inflationary observables, which will be used in the later sections, and the latest observational constraints.

Consider a real scalar field $\phi$ non-minimally coupled to gravity via the Ricci scalar $\mathcal{R}$.
The class of effective actions we consider is
\beq
S=\!\int d^4 x \sqrt{-g}\left[\frac12 \mpl^2 f(\phi)\mathcal{R}-\frac12 \kin(\phi) (\partial \phi)^2-V(\phi)\right],
\label{action}
\eeq
where we allow for a general coefficient of the Ricci scalar $f(\phi)$, general coefficient of kinetic energy $\kin(\phi)$, and general potential $V(\phi)$.
Here $\mpl\simeq 2.43\times 10^{18}$ GeV is the reduced Planck mass; we are effectively assuming that the field $\phi$ is stabilized at the end of inflation with $f(\phi_0)\approx 1$, as will be the case for the Standard Model Higgs.

The cosmology of this theory is most easily studied by performing a conformal transformation to the so-called ``Einstein frame" where the gravity sector is canonical $\frac{1}{2}\mpl^2 \mathcal{R}_E$. This is achieved by defining the Einstein metric as
$g_{\mu\nu}^E=f(\phi)g_{\mu\nu}$. The corresponding Einstein frame potential is
\beq
V_E(\phi)={V(\phi)\over f(\phi)^2}\,.
\eeq
Furthermore, the kinetic energy in the Einstein frame can be made canonical with respect to a new field $\phie$, defined through the equation
\beq
\left({d\phie\over d\phi}\right)^2\equiv
{\kin(\phi)\over f(\phi)}+{3\over 2}\mpl^2 {f'(\phi)^2\over
f(\phi)^2}\,,\label{kinetic}
\eeq
(the second term here comes from transforming the Ricci scalar). In this frame, the action takes the canonical form
\beq
S=\!\int d^4 x \sqrt{-g_E}\left[\frac12 \mpl^2 \mathcal{R}_E-\frac12 (\partial_E \phie)^2-V_E(\phie(\phi))\right],
\label{action}
\eeq
which is amenable to straightforward analysis.

The inflationary dynamics and cosmological predictions is determined by the shape of the potential $V_E$.
In the usual way, we introduce the first and second slow-roll parameters, which control the first and second derivatives of the potential, respectively.  Using the chain rule, these are
\bea
\epsilon(\phi)&=&\frac12 \mpl^2 \left({V_E'\over V_E}\right)^2 
\left({d\phie\over d\phi}\right)^{-2},\label{epsilon}\\
\eta(\phi)&=&\mpl^2\left[
{V_E''\over V_E}\left({d\phie\over d\phi}\right)^{-2}
\!-{V_E'\over V_E}\left({d\phie\over d\phi}\right)^{-3}
\!\left({d^2\phie\over d\phi^2}\right)
\right],\,\,\,\,\,\label{eta}
\eea
where a prime denotes a derivative with respect to $\phi$. 
Similarly, the third slow-roll parameter $\zeta$
is related to the third derivative of the potential
as $\zeta^2=\mpl^4 (d^3V_E/d\phie^3)(dV_E/d\phie)/V_E^2$.

The number of e-foldings of slow-roll inflation is given by an integral over $\phi$: 
\beq
N_e(\phi)={1\over \sqrt{2}\,\mpl}\int_{\phi_{\rm end}}^{\phi}
{d\tilde\phi\over\sqrt{\epsilon(\tilde\phi)}}\left({d\phie\over d\tilde\phi}\right)\,,
\label{Ne}\eeq
where $\phi_{\rm end}$ is the value of the field at the end of inflation, defined by $\epsilon \simeq1$.
The number of e-foldings must be matched to the appropriate normalization of the data set
and the cosmic history, with a typical value being $N_e\simeq 60$; we return to this point in Section \ref{sec:results}.

The amplitude of density perturbations in $k$-space is specified by the power spectrum:
\beq
P_s(k)=\Delta_\mathcal{R}^2\left(\frac{k}{k^*}\right)^{n_s(k)-1},
\eeq 
where $\Delta_\mathcal{R}^2$ is the amplitude at some ``pivot point" $k^*$, predicted by inflation to be
\beq
\Delta_\mathcal{R}^2 = {V_E\over 24\pi^2 \mpl^4\,\epsilon}\Bigg{|}_{k^*},
\label{Delta}\eeq
and measured by WMAP5 to be $\Delta_\mathcal{R}^2 = (2.445\pm 0.096) \times 10^{-9}$ at $k^*=0.002\, \rm{Mpc}^{-1}$ \cite{wmap5}. The corresponding spectral index $n_s$, running of the spectral index $\alpha\equiv dn_s/d\ln k$, and tensor to scalar ratio $r$, are given to good approximation by
\bea
n_s & = & 1-6\epsilon+2\eta\,,\\
\alpha & = & -24\epsilon^2+16\epsilon\eta-2\zeta^2\,,\\
r & = & 16 \epsilon.\label{reqn}
\eea
The combined WMAP5 plus baryon-acoustic-oscillations (BAO) and supernovae (SN) data considerably constrain $n_s$ and $r$. Assuming negligible $\alpha$, as will be the case for running inflation, the constraints are: $0.93<n_s<0.99$ and $r<0.22$ (at 95\% confidence level).


\section{Classical Analysis}
\label{sec:classical}

Without essential loss we can rotate the Higgs doublet so that it
takes the form $H^T=(1/\sqrt{2})(0, v+\phi)$.  Only the real field $\phi$ will play a role in our analysis.
 Specializing to gauge invariant, dimension $\leq 4$ operators, without higher derivatives, the functions $f(\phi)$, $k(\phi)$, and $V(\phi)$ must take the form
\beq
f(\phi)=1+{\xi\phi^2\over \mpl^2},\quad
\kin(\phi)=1,\quad V(\phi)={\lambda\over 4}(\phi^2-v^2)^2\,,
\label{classfns}\eeq
where $v\simeq 246.2$ GeV is the vacuum expectation value for the 
Higgs field, setting the electroweak scale. 
The  self coupling $\lambda$  is in one-to-one correspondence with the Higgs mass, namely $m_h^2=2\lambda v^2$. Current experimental bounds on the Higgs mass (and hence $\lambda$) are as follows:
\bea
114.4\,\mbox{GeV}< & \! m_h\! & \lesssim 182\,\mbox{GeV},\nonumber\\
 0.11< &\! \lambda\! & \lesssim 0.27,
\label{mhbds}
\eea
where the lower bound comes from direct searches and the upper bound comes from a global fit to precision electroweak data  (95\% CL) \cite{PDG}.

In this theory, inflation takes place at energies many orders of magnitude above the electroweak scale 
($\phi^2\ggg v^2$). Hence, during inflation the potential is well approximated by the quartic potential: $V(\phi)=\frac{\lambda}{4}\phi^4$, and this form of the classical potential will be sufficient throughout this Letter. The corresponding potential in the Einstein frame is then
\beq
V_E(\phi)={{\lambda\over 4}\phi^4\over (1+{\xi\phi^2\over \mpl^2})^2}\,,
\eeq
which approaches a constant $V_0\equiv \lambda\mpl^4/4\xi^2$ at large field values $\phi\gg \mpl/\sqrt{\xi}$ (we assume $\xi>0$).
This fact allows slow-roll inflation to take place \cite{Salopek,Kaiser,Bezrukov}. 
It is notable that through this mechanism slow-roll inflation emerges unusually  ``naturally".  

\begin{figure}[t]
\begin{center}
\includegraphics[scale=0.55]{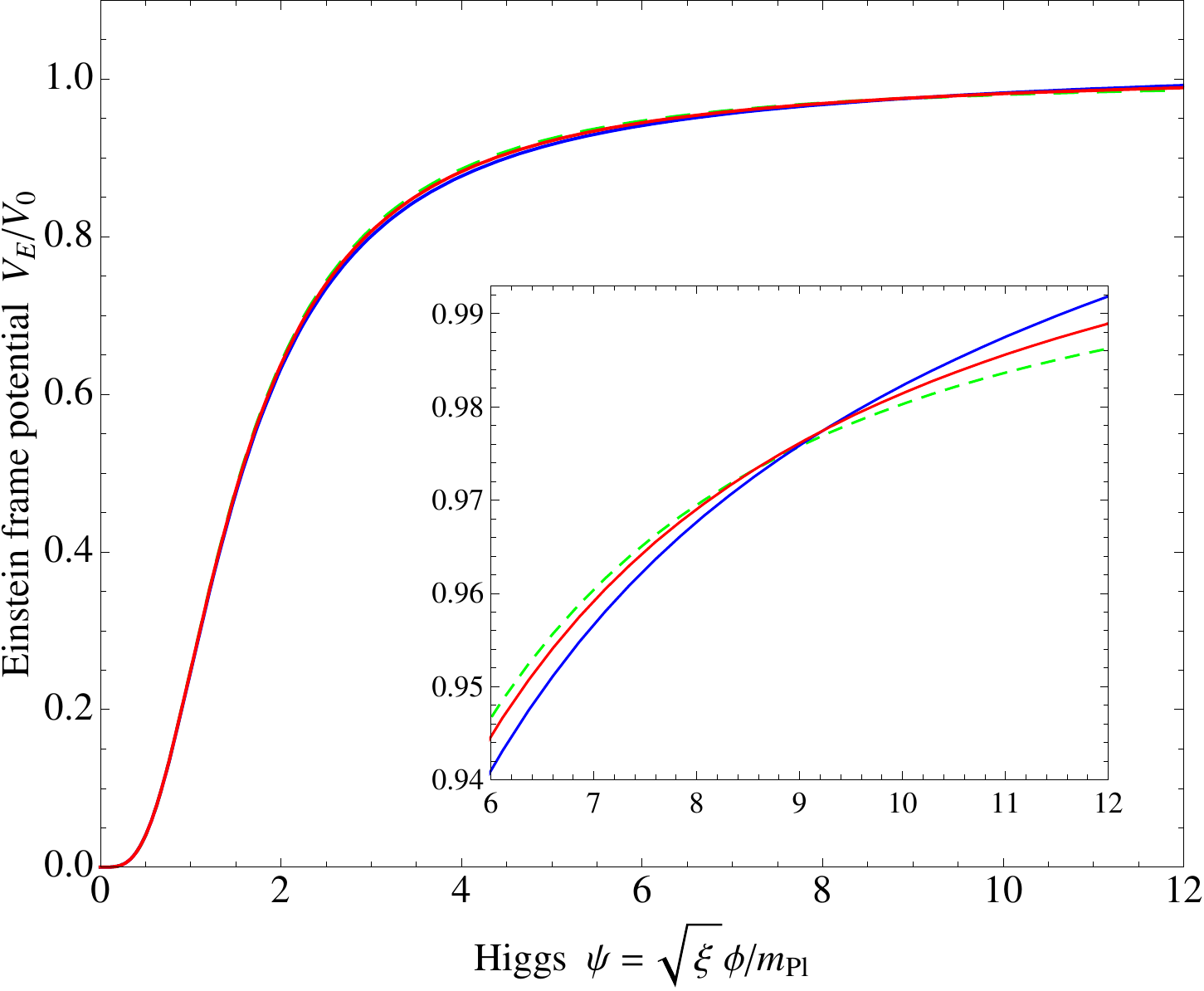}
\end{center}
\caption{The potential in the Einstein frame $V_E$, normalized to a reference value $V_0\equiv \lambda\mpl^4/4\xi^2$, as a function of the Higgs field $\psi=\sqrt{\xi}\,\phi/\mpl$. The dashed green curve is the classical case (independent of Higgs mass), the solid blue (red) curve is the quantum case with Higgs mass $m_h=126.5$\,GeV ($m_h=128$\,GeV). We have set $m_t=171$\,GeV and $\alpha_s(m_Z)=0.1176$ for this plot. The inset focusses on the slow-roll inflationary regime.}
\label{VE}
\end{figure}

It is useful to define the dimensionless quantity  $\phid\equiv\sqrt{\xi}\,\phi/\mpl$
which controls the cosmological evolution: inflationary stage ($\phid\gg1$), 
the end of inflation ($\phid\sim 1$), and the low-energy regime ($\psi\ll 1$). Indeed the potential $V_E$ plotted  in Fig.~\ref{VE} displays the familiar quartic behavior for small $\psi$ values, but asymptotes to a constant for large $\psi$.

Using eqs.~(\ref{kinetic}), (\ref{epsilon}), and (\ref{eta}),
the slow-roll parameters are readily computed.  The exact results are not very transparent.  They simplify for
large $\xi$, which is the case of physical interest:
\beq
\epsilon\simeq {4\over 3\phid^4},\quad
\eta\simeq -{4\over 3\phid^2}\left(1-{1\over \phid^2}\right),\quad
 \zeta^2\simeq {16\over 9 \phid^4}\left(1-{3\over\phid^2}\right)\,.\label{csr}
\eeq
We see that at large $\phid$ (during slow-roll inflation) $\eta$ is dominant, and will primarily control the predictions for the spectral index. The number of e-foldings is computed from eq.~(\ref{Ne}) giving
\beq
N_e\simeq {3\over 4}\left[\phid^2-\phid_{\rm end}^2-
\ln\left({1+\phid^2\over 1+\phid_{\rm end}^2}\right)
\right]\,,\label{cNe}
\eeq
where $\phid_{\rm end}\simeq(4/3)^{1/4}$ is the value of $\phid$ at the 
end of inflation ($\epsilon\simeq1$).
Eqs.~(\ref{csr}) and (\ref{cNe}) provide a parametric description of $\epsilon(N_e)$, $\eta(N_e)$, and $\zeta(N_e)$, thus determining $n_s$, $\alpha$, and $r$ as a function of $N_e$, i.e., we can trade the unknown value of the Higgs field during inflation $\phi(=\mpl\phid/\sqrt{\xi})$ for the number of e-foldings $N_e$. 

For $N_e = 60$ we find the following results for the spectral index, the running of the spectral index, and the tensor to scalar ratio:
\beq
n_s\simeq 0.968, \quad
\alpha\simeq -5.2\times 10^{-4},\quad
r\approx 3.0\times 10^{-3}.
\label{nsclass}
\eeq
We see that $\alpha$ and $r$ are rather small. This will remain qualitatively true in the quantum theory, but the corrections to $n_s$ are quite important, as we explore in detail in the next section.

Finally, using eq.~(\ref{Delta}) and expanding to leading order in $1/\psi\sim1/\sqrt{N_e}$, the amplitude of density fluctuations is found to be
\beq
\Delta_{\mathcal{R}}^2 \simeq {\lambda\over\xi^2}{N_e^2\over 72\pi^2}.
\eeq
Since this must be $\mathcal{O}(10^{-9})$, it is impossible to satisfy for $\lambda=\mathcal{O}(0.1)$ and $\xi=\mathcal{O}(0.1)$ (which might be considered ``natural'' values). One possibility is that $\lambda$ is extremely small, but that is incompatible with experimental bounds on the Higgs mass, see eq.~(\ref{mhbds}), and is not stable under renormalization. Instead, following \cite{Bezrukov}, we assume $\xi=\mathcal{O}(10^4)$ in order to obtain the correct amplitude of density fluctuations with $\lambda=\mathcal{O}(0.1)$. The need to dial a parameter to large or small values, so that $\Delta_{\mathcal{R}}^2$ is consistent with observations, is a common feature to all known inflation models.  It will also apply in the quantum theory.

\section{Quantum Analysis}
\label{sec:quantum}
We now consider how quantum corrections modify the classical results of the previous section. 
In order to do so, we need to compute the effective action that takes into account the effects of particles of the Standard Model interacting with the Higgs boson through quantum loops. 
The frame we calculate in is the original ``Jordan" frame which defines the theory.
The quantum theory modifies all three functions $f(\phi)$, $\kin(\phi)$, $V(\phi)$ from the classical expressions in eq.~(\ref{classfns}). 


The quantum corrections to the classical kinetic sector $\kin(\phi)=1$ arise from wave-function renormalization, and are approximately $\xi$--independent. It is simple to check that at large $\xi$ the second term in eq.~(\ref{kinetic}) scales as $\xi^0\sim 1$, while the contribution from the $\kin(\phi)/f(\phi)$ term scales as $1/\xi$. Hence corrections to $\kin(\phi)$ occur with a factor $1/\xi$, in addition to suppression by loop factors and couplings.

\begin{figure}[t]
\begin{center}
\includegraphics[scale=0.75]{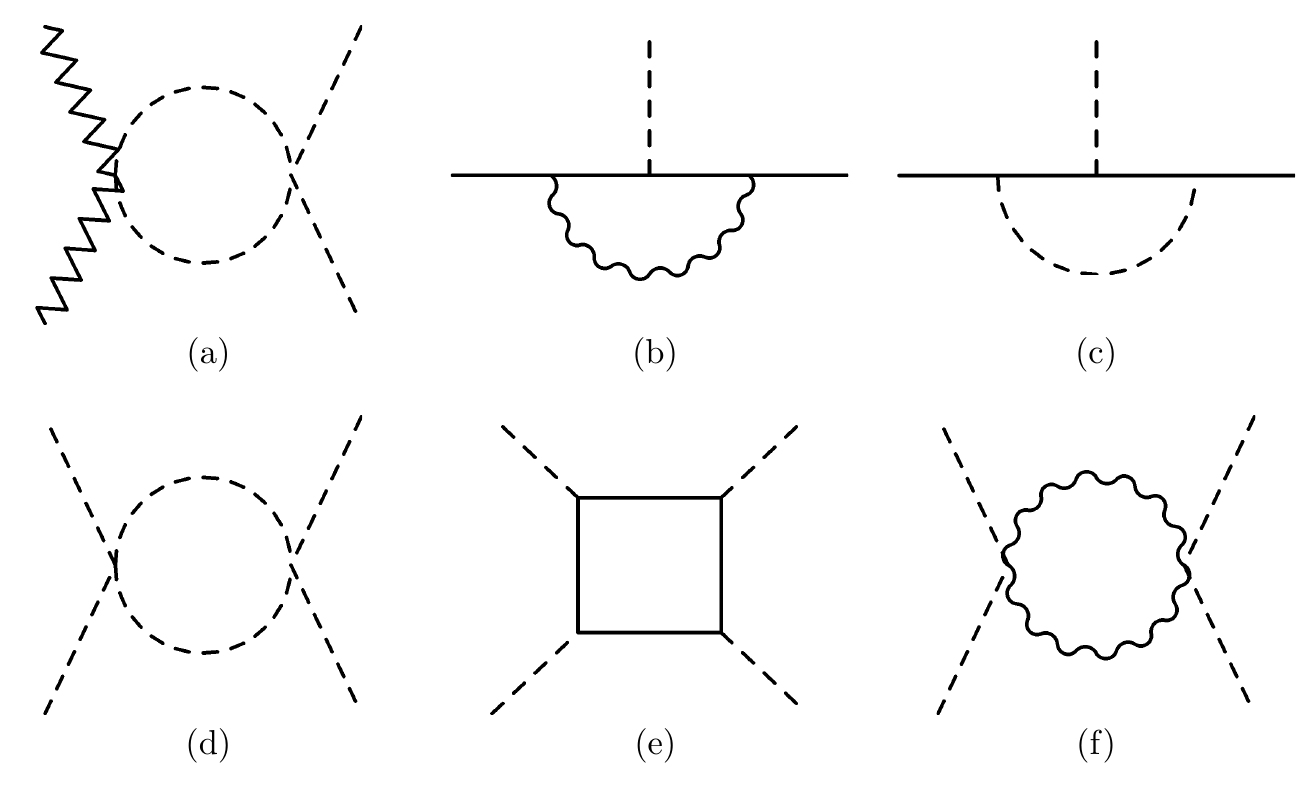}
\end{center}
\caption{Some representative Feynman diagrams. 
Top row: renormalization of the conformal coupling $\xi$ with Higgs in loop (a), and renormalization of top quark's Yukawa coupling with gauge boson (b) and  Higgs (c) across vertex.
Bottom row: renormalization of quartic coupling $\lambda$ with Higgs (d), top quark (e), and 
gauge boson (f) in loop. }
\label{diagrams}
\end{figure}

The quantum corrections to the classical gravity sector $f(\phi)=1+\xi\phi^2/\mpl^2$ are more subtle.
Let us start by considering the case of a (classical) {\em background} gravitational field. In this case the conformal anomaly induces a 1-loop $\beta$-function for $\xi$ given by \cite{confbeta}
\beq
\beta_\xi = {6\xi+1\over(4\pi)^2}\left[2\lambda+y_t^2-{3\over 4} g^2-{1\over 4}g'^2\right].
\eeq
The term proportional to $\lambda$, coming from Higgs running in a loop (see Fig.~\ref{diagrams}(a)), is potentially important during inflation. We will return to this point soon when we include the (classical) {\em back reaction} of gravity, and argue that in fact this contribution is negligible. The remaining terms arise from external leg corrections and cancel against wave-function renormalization to good approximation. Hence corrections to $f(\phi)$ are ignorable also.

\begin{figure}[t]
\begin{center}
\includegraphics[scale=0.285]{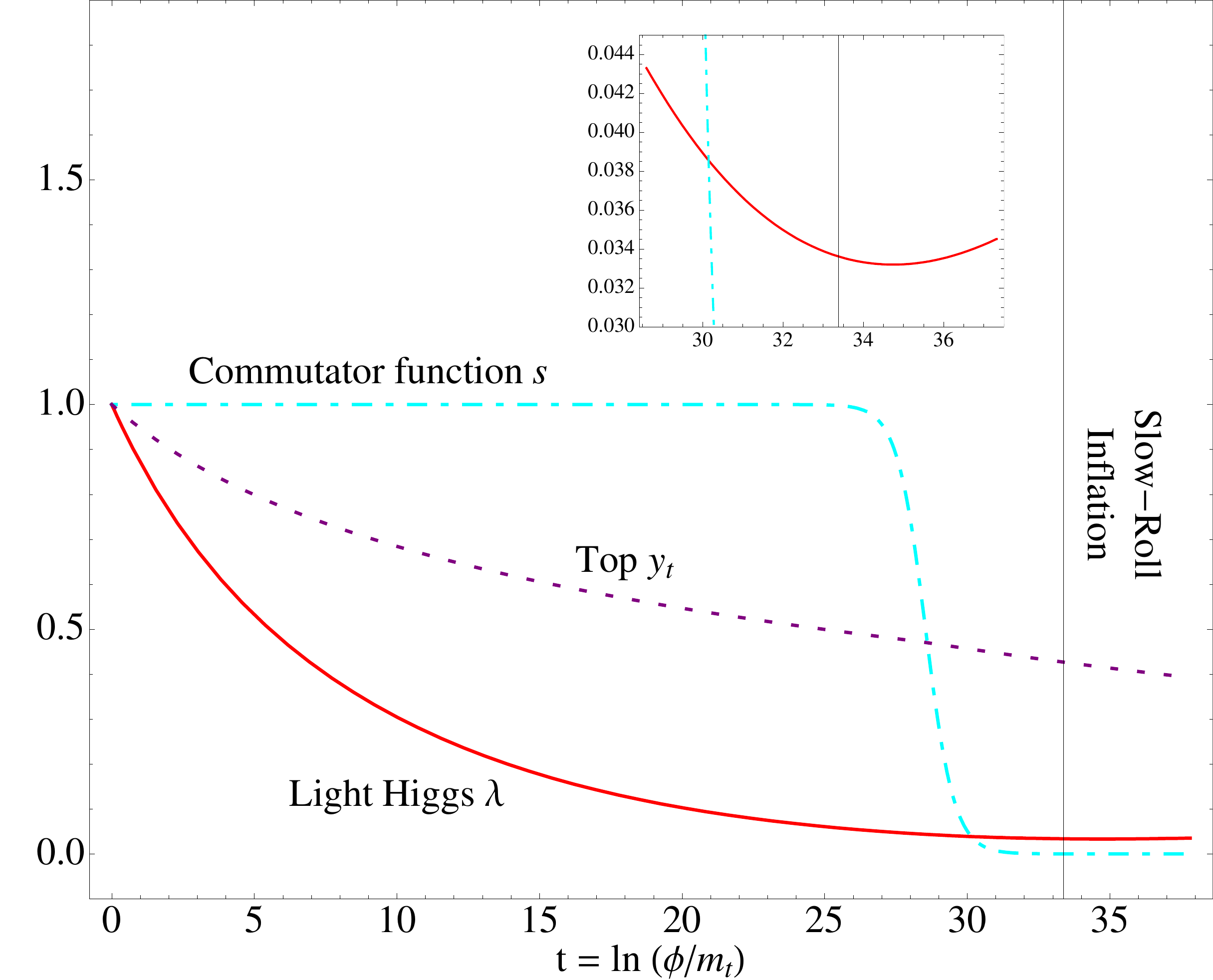}\,\,\,\,\,
\end{center}
\caption{This plot summarizes some of the most important effects of the renormalization group flow.
The red curve shows the running of the quartic coupling $\lambda(t)/\lambda(0)$ for a light Higgs $m_h=126.5$\,GeV. The dotted purple curve is the top running $y_t(t)/y_t(0)$ and the dot-dashed cyan curve is the commutator function $s(t)$, with $\xi= 2.3\times 10^3$ and $\mu=m_t$. The right-hand region is the slow-roll inflationary regime; here $\lambda$ rises (and so $n_s$ does too), as highlighted by the inset. }
\label{runplot}
\end{figure}

Finally we turn to the computation of the potential sector $V(\phi)$. Let us begin with the flat space analysis. The RG improved potential for the Higgs in the Standard Model is 
(see e.g.~Ref.~\cite{Sher} for a review)
\beq
V(\phi)={1\over 4}\lambda(t(\phi))\,G(t(\phi))^4\,\phi^4,
\label{RGimp}\eeq
($\phi\ggg v$) where $t(\phi)=\ln(\phi/\mu)$, and $\mu$ is the normalization point; taken to be $\mu=m_t$ in this Letter.
Here $\lambda(t)$ encodes the running of $\lambda$, while 
$G(t)=\exp(-\int_0^t dt' \gamma(t')/(1+\gamma(t')))$,
where $\gamma$ is the anomalous dimension of the Higgs field, encodes wave-function renormalization. The running of $\lambda$ is governed by the renormalization group equation: $d\lambda/dt=\beta_{\lambda}/(1+\gamma)$. At 1-loop it is
\bea
\beta_\lambda={1\over(4\pi)^2}
\Bigg{[}24\lambda^2-6\,y_t^4+{9\over 8}g^4+{3\over 4}g^2g'^2+{3\over 8}g'^4
+\,\lambda(12y_t^2-9g^2-3g'^2)\Bigg{]}.
\label{betalam}\eea
At low energies, the two most important terms here are the self coupling $24\lambda^2$ (see Fig.~\ref{diagrams}(d)), which tries to drive $\lambda$ to large positive values, and the top quark $-6\,y_t^4$ (see Fig.~\ref{diagrams}(e)), which tries to drive $\lambda$ towards zero. This is summarized in Fig.~\ref{runplot}.
This leads to a delicate interplay between the Higgs mass and the top mass.
For $m_h\gg m_t$, the $24\lambda^2$ term dominates and $\lambda$ will eventually hit a Landau pole at high energies. 
For $m_h\ll m_t$, the $-6\,y_t^4$ dominates and $\lambda$ will go negative which is a sign of vacuum instability.
The ``Goldilocks" window for the Higgs mass, where the theory is both perturbative and stable up to very high energies is also the regime in which the quantum corrections are relatively small, allowing for slow-roll inflation. At high energies, the contribution from gauge bosons (see Fig.~\ref{diagrams}(f)) are important and increase $\lambda$.

In the recent work of Barvinsky \textit{et al.}~\cite{Barvinsky} the top quark's Yukawa coupling was approximated by the tree level value: $y_t=\sqrt{2}\,m_t/v$ for all energy scales. This provides a significant negative contribution to $\beta_\lambda$, forcing $\lambda$ to negative values and vacuum instability in large regions of parameter space. Instead it is essential to include the running of the top Yukawa coupling in the analysis:
\beq
\beta_{y_t} = {y_t\over(4\pi)^2}\left[{9\over 2}y_t^2 -8g_s^2-{9\over 4}g^2-{17\over 12}g'^2\right],
\eeq
which is negative due to the large negative contribution from the strong coupling $-8g_s^2 y_t$ (see Fig.~\ref{diagrams}(b)).
Hence $y_t$ runs to smaller values at high energies; see Fig.~\ref{runplot}.

In our work, we have included the complete running of the 5 couplings: $\lambda$, $y_t$, $g_s$, $g$, and $g'$ to 2 loops, to ensure accurate results.\footnote{The 3-loop running is unknown for the Standard Model, but would need to be abnormally large to have an effect.}
The $\beta$-functions are summarized in Appendix \ref{App}. Furthermore, we have adopted the pole mass matching scheme for the Higgs and top masses, given in the Appendix of \cite{Giudice}. For the sake of brevity, we do not reproduce the pole matching details here.


We now consider the effective potential $V(\phi)$ including the effect of the non-minimal coupling to gravity $\xi\phi^2\mathcal{R}$. The calculation is difficult to perform exactly. However, we can obtain 
approximate results for large $\xi$ fairly simply. Following \cite{Salopek}, one can heuristically identify a non-standard commutator for $\phi$ as follows. 
From eqs.~(\ref{kinetic}) and (\ref{action}) we see that when the gravity sector is canonical, the kinetic sector is non-canonical
$-{1\over 2}(\partial_E\phi)^2\left(d\phie/d\phi\right)^2.$
On a spatial hypersurface, the canonical momentum corresponding to $\phi$ is 
\bea
\pi={\partial\mathcal{L}\over\partial\dot{\phi}}
=\sqrt{-g_E}\left(g_{E}^{\mu\nu}\,n_\mu\, \partial_\nu\phi\right) \left({d\phie\over d\phi}\right)^2
=\sqrt{-g}\left(g^{\mu\nu}\,n_\mu \, \partial_\nu\phi\right) f(\phi)\left({d\phie\over d\phi}\right)^2,
\eea 
where $n_\mu$ is a unit timelike vector.
Imposing standard commutation relations for $\phi$ and $\pi$, we learn that
 $[\phi({\bf x}),\dot{\phi}({\bf y})] = i\,\hbar\,\sfun(\phi)\,\delta^{(3)}({\bf x}-{\bf y})$,
with 
\beq
\sfun(\phi) = f^{-1}(\phi)\left({d\phie\over d\phi}\right)^{-2} =
{1+{\xi\phi^2\over\mpl^2}\over 1+(6\xi+1){\xi\phi^2\over\mpl^2}}.
\label{comm}\eeq
For $\phi\ll \mpl/\xi$ (the low energy regime) we recover the ordinary value of the commutator $s=1$, while for $\phi\gg \mpl/\sqrt{\xi}$ (the inflationary regime) we see a suppression in the commutator by a factor of $s=1/(6\xi+1)$. So in the inflationary regime with $\xi\gg 1$, quantum loops involving the Higgs field are heavily suppressed. 

To summarize, our prescription for the renormalization group improved effective potential in the presence of non-minimal coupling is to assign one factor of $\sfun(\phi)=\sfun(\mu\,e^t)$ for every off-shell Higgs that runs in a  quantum loop. This factor is plotted as the dot-dashed cyan curve in Fig.~\ref{runplot}. In eq.~(\ref{betalam}), for example, this prescription means the replacement $24\,\lambda^2\to24\, s^2\,\lambda^2$, as that term arises from two Higgs off-shell propagators, while all other terms are untouched since they only involve other fields in loops (see Appendix \ref{App} for more details).
This provides an important modification to the high energy running of couplings,
and explains why the running of $\xi$ from the diagram of Fig.~\ref{diagrams}(a) is suppressed.
Apart from this modification, the RG improved analysis is as standard, as summarized in eq.~(\ref{RGimp}). 
We have checked our prescription against detailed analytical calculations of the effective action of non-minimally coupled scalars in the literature (e.g., see \cite{Barvinsky2,Barvinsky3}) and have found excellent agreement. We assume that quantum corrections from graviton exchange are small, see Appendix \ref{NewApp}.

\section{Results and Predictions}
\label{sec:results}
After numerically solving the set of 5 coupled renormalization group differential equations of Appendix \ref{App} for the couplings: $\lambda$, $y_t$, $g_s$, $g$, and $g'$, we have obtained the effective potential $V(\phi)$ in the full quantum theory, as a function of input parameters, such as the Higgs mass. Some representative potentials in the Einstein frame are given in Fig.~(\ref{VE}). The inset clearly exhibits variation of the effective potential $V_E(\phi)$ with Higgs mass, which was absent in classical case. As we lower the Higgs mass,  approaching the instability, the magnitude of the first derivative is raised and the that of the second is lowered (see the blue and red curves). 
This leads to modifications to the cosmological parameters.

Following the recipe we outlined earlier in Section \ref{sec:nonmin},
we are able to efficiently compute the spectral index in the RG improved theory using Mathematica.
Recall that in the classical theory, $n_s$ is independent of the parameters of the 
Standard Model, and its value was found to be $n_s\simeq 0.968$ (for $N_e=60$).
In the quantum theory, we find that $n_s$ depends on several of the Standard Model parameters,
in particular on  the Higgs and top masses, see Fig.~\ref{nsplot}.
As the top mass is varied through its experimentally allowed range ($169\,\mbox{GeV}\lesssim m_t\lesssim173\,$GeV) the spectral index varies noticeably.
In particular, as we lower the Higgs mass towards vacuum instability, the spectral index increases substantially.
To achieve successful inflation with $n_s < 0.99$, we require
\beq
m_h>125.7\,\mbox{GeV} + 3.8\,\mbox{GeV}\left({m_t-171\,\mbox{GeV} \over 2\,\mbox{GeV}}\right)
-1.4\,\mbox{GeV}\left({\alpha_s(m_Z)-0.1176\over 0.0020}\right)\pm \delta,
\label{boundsmh}\eeq
where $\delta\sim 2$\,GeV indicates theoretical uncertainty from higher order corrections 
(such as 3-loop). 
This bound almost coincides with that from absolute stability presented in Ref.~\cite{Giudice}.
Note that near the boundary $\lambda$ is small, so the corresponding $\xi$ to obtain the observed 
$\Delta_{\mathcal{R}}^2$ is reduced from its classical value $\xi\sim 10^4$ by an order of magnitude or so to $\xi\sim 10^3$.

Let us now trace the chain of logic behind the rise in $n_s$. For a light Higgs, $\beta_\lambda$ is dominated by the top and gauge boson contributions. For a heavy top, the top contribution is dominant at low energies, causing $\beta_\lambda$ to be negative and thus driving $\lambda$ to low values as the energy is increased. At the same time, the top Yukawa coupling runs, with dominant contributions coming from gauge fields and Higgs running in a loop, with the gauge fields slightly dominant causing $y_t$ to decrease with energy.\footnote{Note that the 2-loop term $-108y_tg_s^4/(4\pi)^4$ in $\beta_{y_t}$ (see eq.~(\ref{betayt})) speeds up the running compared to 1-loop.}
At very high energies $\phi\gg \mpl/\sqrt{\xi}$ (the inflationary regime), the Higgs running in the loop is highly suppressed, causing $y_t$ to jump to even lower values. Hence the top contribution to the running of $\lambda$ becomes subdominant, the gauge boson contributions now dominate and $\lambda$ rises, as seen in Fig.~\ref{runplot} (inset). Since $\lambda$ is concave up, this increases $\eta$ and hence the spectral index.

\begin{figure}[t]
\begin{center}
\includegraphics[scale=0.78]{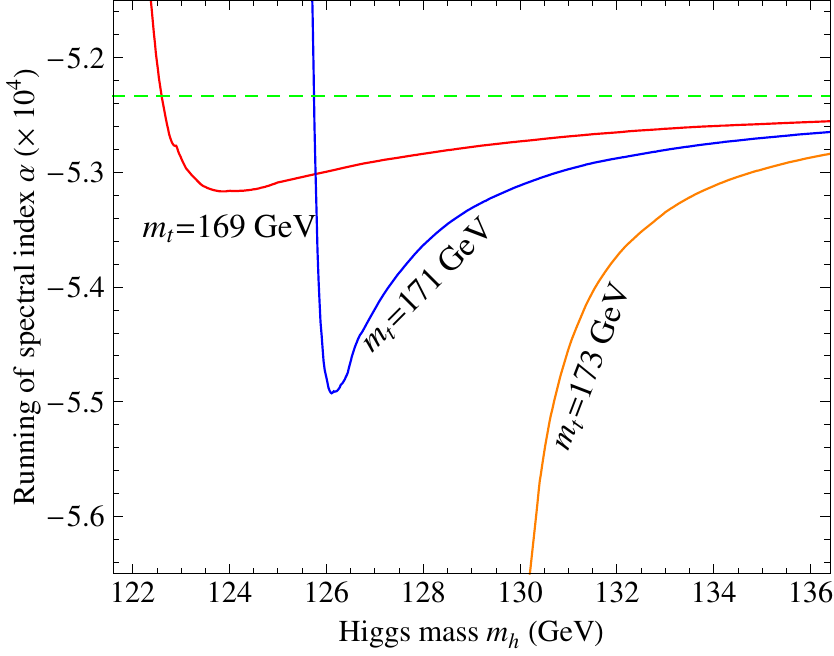}\,\,\,\,
\includegraphics[scale=0.77]{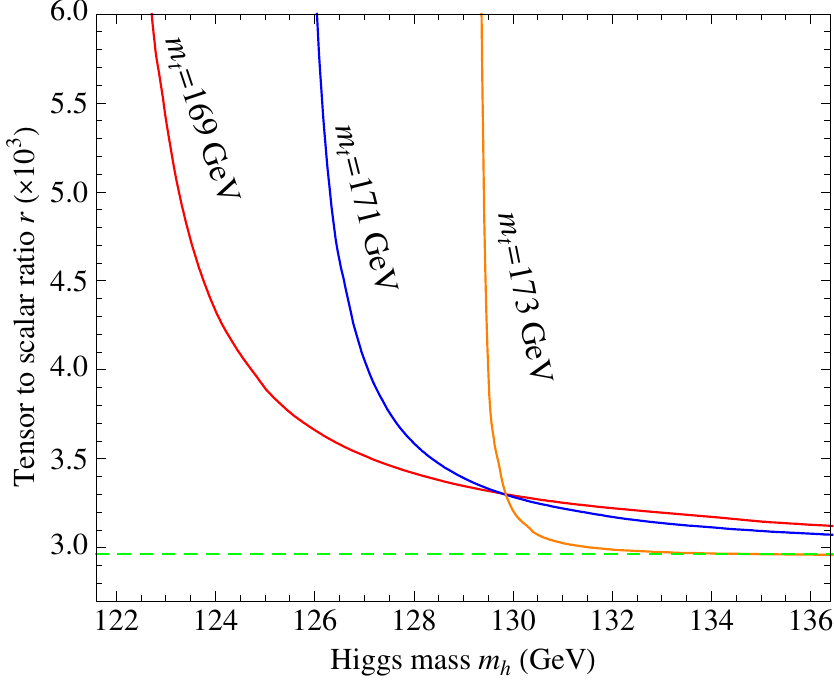}
\caption{The running of the spectral index $\alpha\,(\times 10^4)$ (left panel) and the tensor to scalar ratio 
$r\,(\times 10^3)$ (right panel) as a function of the Higgs mass $m_h$.
The 3 solid curves correspond to 3 different values of the top mass: $m_t=169$\,GeV (red curve), $m_t=171$\,GeV (blue curve), and $m_t=173$\,GeV (orange curve). 
The horizontal dashed green curve, with $\alpha\simeq -5.2\times 10^{-4}$ and $r\simeq 3.0\times 10^{-3}$, is the classical result. We have set $\alpha_s=0.1176$ and $N_e=60$ in this plot.}
\end{center}
\label{rplot}
\end{figure}

In Fig.~\ref{nsplot} and in all plots we have chosen the reference value $N_0=60$.
For $N_e$ close to $N_0$, we can Taylor expand $n_s$ to linear order:
\beq
n_s(N_e) = n_s(N_0)+{d n_s\over dN_e}(N_e-N_0)+\ldots.
\eeq
Now, the spectral index is in fact a function of all the parameters, including $N_e$ and $\xi$: $n_s=n_s(N_e,\xi,\ldots)$. As in the classical theory, we have fixed $\xi$ such that the amplitude of density fluctuations is in agreement with observations (requiring $\xi\sim 10^4\sqrt{\lambda}$). In this way, we can think of $\xi=\xi(N_e)$, so from the chain rule
\beq
{dn_s\over dN_e}={\partial n_s\over\partial N_e}+{\partial n_s\over\partial\xi}{d\xi\over dN_e}.
\eeq
The first term is precisely the (negative) of the running of the spectral index $\alpha= d n_s/d\ln k$, while the second term is found to be very small numerically. Hence to a good approximation we can write
\beq
n_s(N_e)\approx n_s(N_0)-\alpha(N_0)(N_e-N_0).
\label{nsapprox}\eeq
We plot the running of the spectral index $\alpha(N_0=60)$ in Fig.~\ref{rplot} (left). We see that $\alpha\approx -5\times 10^{-4}$ (as in the classical case), with some variation for low Higgs masses as we approach the instability. However, this is still far too small to be detected by PLANCK, which is expected to be only sensitive to $\alpha=\mathcal{O}(10^{-2})$ \cite{Liddlerunning}. Hence the main usefulness of Fig.~\ref{rplot} (left) is that it should be used in accompaniment with Fig.~\ref{nsplot} and eq.~(\ref{nsapprox}) to infer the value of $n_s$ for different values $N_e$ (as long as $N_e$ does not vary too far from $N_0=60$).

The actual number of e-foldings of inflation is related to the wavenumber of interest $k$, the energy density during inflation $V_E$, the energy density at the end of inflation $V_{\mbox{\tiny{end}}}$, and the energy density at the end of reheating
$\rho_{\mbox{\tiny{reh}}}$ \cite{ModLiddle}
\bea
N_e\simeq 62-\ln{k\over a_0 H_0} + {1\over 4}\ln{V_E \over (10^{16}\,\mbox{GeV})^4}+{1\over 4}\ln{V_E\over V_{\mbox{\tiny{end}}}}-{1\over 12}\ln{V_{\mbox{\tiny{end}}}\over\rho_{\mbox{\tiny{reh}}}}.\label{Neformula}
\eea
Since the Higgs is strongly coupled to Standard Model fields, reheating is expected to occur automatically.
As $N_e$ has only a weak dependence on $\rho_{\mbox{\tiny{reh}}}$, the details of reheating are rather inconsequential to our mass bounds, but may be calculable \cite{Bellido}.
According to \cite{Bezrukov2}, $T_{\mbox{\tiny{reh}}}\sim 10^{13.5}$\,GeV giving $N_e\simeq 59$ for the classical theory.
In our case, we must take into account the variation in the scale of inflation due to the quantum corrections. In Fig.~\ref{rplot} (right) we plot $r$ versus the Higgs mass. Since we have fixed $\xi$ such that the amplitude of density fluctuations is at the observed value, the energy density of inflation $V_E$  is simply proportional to $r$. Using eqs.~(\ref{Delta}) and (\ref{reqn}), we have
\beq
V_E = {3\over 2}\pi^2 \mpl^4\,\Delta_{\mathcal{R}}^2\, r \approx \left(7.8\times 10^{15}\,\mbox{GeV}\right)^4 \left({r\over 0.003}\right).
\eeq
 Since $r$ changes by a factor of order 2, as we vary the Higgs mass, then $V_E$ changes by the same amount. From eq.~(\ref{Neformula}), rescaling $V_E$ and $V_{\mbox{\tiny{end}}}$ by a factor of $p$, say, the number of e-foldings is shifted by $\Delta N_e={1\over 6}\log p$, which is $\approx 0.1$ for $p =2$. Hence the variation in $N_e$ with the Higgs mass is very small.


\section{Discussion}
\label{sec:discussion}
A number of papers have discussed bounds on the Higgs mass coming from demanding stability of the vacuum, e.g., see \cite{Sher,Giudice,Nimabds,Isidori}. Cosmological constraints only require metastability on the lifetime of the universe, which places the constraint $m_h\gtrsim 105$\,GeV \cite{Giudice}. However, if we further demand that the Higgs drive inflation, we find that heavier Higgs are required: $m_h\gtrsim 126$\,GeV (depending on the top mass, see eq.~(\ref{boundsmh})), which essentially coincides with the bounds from absolute stability. Furthermore, by demanding that the theory remains perturbative to high energies ($m_h\lesssim 190$\,GeV),
we establish a correlation between both stability and triviality bounds, and inflation.

More precisely, {\em we have established a  mapping between the renormalization group flow and the cosmological spectral index}.  Over a substantial range of parameter space the classical value $n_s\simeq 0.968$ (for $N_e=60$) emerges as a good approximation, but there are corrections.  Given a detailed microphysical theory, such as the Standard Model, we can explicitly calculate such corrections, as summarized in Fig.~\ref{nsplot}. This plot displays a sharp rise in the spectral index towards 0.98, or so, as we approach vacuum instability for a light Higgs.


It is likely that the Standard Model  is only  the low-energy limit of a more complete
theory,  accommodating the facts that do not find explanation within
the Standard Model, such as neutrino masses, dark matter, baryon asymmetry, etc.
Our methodology is still applicable, so long as we can control the relevant $\beta$-functions.

In principle some quite different scalar field, not connected to the Standard Model Higgs, could drive running inflation.   The central requirement is a large coefficient for the $\phi^2  \mathcal{R}$ term.   It is possible that such a coefficient could emerge as some sort of Clebsch-Gordan coefficient, or from the coherent addition of several smaller terms (involving more basic scalars $\phi_j$).   It is also possible to consider, in the same spirit, the dimension 3 interaction  $\phi  \mathcal{R}$, which arises for generic scalar fields, though not of course for the Standard Model Higgs. 
Furthermore, as discussed in Appendix \ref{NewApp} and Refs.~\cite{Burgess,Espinosa}, the inclusion of higher dimension operators may significantly affect the predictions of the original theory and even spoil its validity (as is the case in many inflationary models).
  These possibilities, and their possible embedding in unified field theory or string theory, deserve further investigation.

If the Higgs boson exists and it is in the  mass  range considered in this Letter, 
the LHC will discover it and will determine
the Higgs mass with a precision of about $0.1\%$ \cite{tdr}, which means
an uncertainty $\Delta m_h\approx 0.2$ GeV.
In order to extract accurate correlations between the inflationary
observables and the Higgs mass it is crucial to improve the precision
with which we know the other parameters of the Standard Model, 
in particular the top quark mass and the strong coupling.
The current value of the top mass from direct observation of events is
$m_t=(171.2\pm 2.1)$ GeV \cite{PDG}. In the near future,
the LHC will improve the determination of the top mass, but relatively 
large systematic uncertainties will prevent a top mass determination to better
than $1$ GeV; more conservatively, the top mass will be determined at LHC
with an error $\Delta m_t=1 \div  2$ GeV.
Looking further ahead, the ILC is expected to be able to measure the top mass to $\sim100$\,MeV. 
So together with the measured Higgs mass from the LHC and improved precision on the strong coupling, as well as calculating higher order effects and reheating details, running inflation in the Standard Model will predict a rather precise value for the spectral index.

\bigskip

We would like to thank M.~Amin, F.~D'~Eramo, A.~Guth, and C.~Santana for useful discussions,
and A.~Riotto for comments on the manuscript. We thank F.~L.~Bezrukov, A.~Magnin, and M.~Shaposhnikov for correspondence.
The work of ADS is supported 
in part by the INFN ``Bruno Rossi'' Fellowship. The work of ADS, MPH, and FW is supported in part by the U.S. Department of Energy (DoE) under contract No. DE-FG02-05ER41360. 


\section*{Note Added}
Our Letter appeared simultaneously on the arXiv with Ref.~\cite{Bez2008}, which also studied the quantum corrections to inflation driven by the Standard Model Higgs. 
The central conclusion of both papers is that the classical analysis provides a good approximation over a wide range of parameters, but that  quantum corrections are calculable and can be quantitatively significant.

For a top mass of $m_t=171.2$\,GeV,  Ref.~\cite{Bez2008} found that in order to have successful inflation the Higgs mass is constrained to be in the range: $136.7\,\mbox{GeV}<m_h<184.5\,\mbox{GeV}$, and the spectral index decreases from its classical value as $m_h$ approaches the lower boundary. In this note we briefly discuss the similarities and differences between their analysis and ours.\footnote{The discussion here refers to version 1 of \cite{Bez2008}. 
}

In our analysis, we computed the full RG improved effective potential. We did this including (i) 2-loop beta functions, (ii) the effect of curvature in the RG equations (through the function $s$), (iii) wave-function renormalization, and (iv) accurate specification of the initial conditions through proper pole matching. On the other hand, \cite{Bez2008} did not compute the full effective potential or include any of the items (i)--(iv).\footnote{Though wave-function renormalization was not included in \cite{Bez2008}, external leg corrections in the running of $\lambda$ were included. However these two effects roughly cancel against one another.} Instead Ref.~\cite{Bez2008} approximated the potential at leading log order with couplings evaluated at an inflationary scale after running them at 1-loop (this is one step beyond \cite{Barvinsky} where couplings were not run).

The lower bound on the Higgs mass we find in eq.~(\ref{boundsmh}) is about 11\,GeV lower than that found in Ref.~\cite{Bez2008} ($m_h>136.7$\,GeV). This numerical discrepancy is due to several of the above simplifications, but the dominant difference comes from (i) inclusion or not of  2-loop effects (importantly, the $-108y_tg_s^4/(4\pi)^4$ term in $\beta_{y_t}$, see eq.~(\ref{betayt})), and a second significant difference comes from (iv) pole matching. Higher order effects (such as 3-loop) and uncertainty in the strong coupling $\alpha_s(m_Z)$ also modify the bound, as we summarized in eq.~(\ref{boundsmh}).

A precise upper bound on the Higgs mass  ($m_h<184.5$\,GeV) is stated in \cite{Bez2008}.  The basis of this is the famous ``triviality bound'', see e.g., Ref.~\cite{Ellis}, which has little to do with inflation.    The theory ultimately requires a cutoff, and exactly how low a cutoff one feels comfortable with (or equivalently, how large a value of $\lambda$ one regards as acceptable) is arguable. We feel that our stated semi-quantitative bound $m_h\lesssim 190$\,GeV adequately represents the situation.

In \cite{Bez2008} $n_s$ decreases as $m_h$ approaches its minimum value, while we find that $n_s$ increases (see Fig.~\ref{nsplot}). This behavior depends critically on the value of $y_t$ during inflation, as compared to the value of the gauge couplings. If $y_t$ is small, then $n_s$ increases, and vice versa.
Ref.~\cite{Bez2008} overestimated $y_t$ during inflation and hence obtained the opposite behavior. This is primarily due to ignoring items (i) and (ii) above. Ignoring (i) misses the 2-loop term $-108y_tg_s^4/(4\pi)^4$ in $\beta_{y_t}$, and ignoring (ii) maintains the 1-loop term $\frac{9}{2}y_t^3/(4\pi)^2$ in $\beta_{y_t}$ during inflation.

Finally, \cite{Bez2008} computes quantum corrections with both a field-independent cutoff (as we use) and a field-dependent cutoff in the original ``Jordan" frame.   Either procedure defines a possible model, but the field-independent cutoff is more in the spirit of the motivating arguments, based on dimension $\leq 4$ effective Lagrangians.  


\appendix
\section{2-Loop RG Equations}
\label{App}

In this appendix we list the RG equations for the couplings $\lambda,\,y_t, g',g,g_s$  at energies above $m_t$ at 2-loop \cite{Ford}. In each case, we write $d\lambda/dt=\beta_\lambda/(1+\gamma)$, etc., where $t=\ln \phi/\mu$. Also, we insert one factor of the commutator function $s(\mu\,e^t)$ (see eq.~(\ref{comm})) for each off-shell Higgs propagator.\footnote{We have carefully extracted out all Higgs propagators contributions at 1-loop order by the appropriate insertion of factors of $s$. For the 2-loop contributions we have only inserted $s$ for the obvious terms. The complete set of insertions are tedious and provide negligible corrections.} 

For the Higgs quartic coupling we have
\bea
\beta_\lambda&=&
 {1\over(4\pi)^2}  \left[24 s^2 \lambda ^2-6 y_t^4+\frac{3}{8} \left(2 g^4+\left(g^2+g'^2\right)^2\right)+\left(-9 g^2-3
   g'^2+12 y_t^2\right) \lambda \right]\nonumber\\
   &+&{1\over (4\pi)^4}
  \Bigg{[}\frac{1}{48} \left(915 g^6-289 g^4 g'^2-559 g^2 g'^4-379 g'^6\right)+30
   s y_t^6-y_t^4 \left(\frac{8 g'^2}{3}+32 g_s^2+3 s \lambda
   \right)\nonumber\\
   &+& \lambda  \left(-\frac{73}{8} g^4+\frac{39}{4} g^2 g'^2+\frac{629
   }{24}s g'^4+108 s^2 g^2  \lambda +36s^2 g'^2 \lambda -312
   s^4 \lambda ^2\right)\nonumber\\
   &+& y_t^2 \left(-\frac{9}{4} g^4+\frac{21}{2} g^2
   g'^2-\frac{19}{4}g'^4+ \lambda  \left(\frac{45}{2}g^2+\frac{85
   }{6}g'^2+80 g_s^2-144 s^2 \lambda \right)\right)\Bigg{]}.
 \eea

For the top Yukawa coupling we have
\bea
\beta_{y_t} &=&
 {y_t\over(4\pi)^2} \left[-\frac{9}{4} g^2-\frac{17
   }{12}g'^2-8 g_s^2+\frac{9}{2} s
   y_t^2\right]
+{y_t\over(4\pi)^4}
   \Bigg{[}-\frac{23}{4} g^4-\frac{3}{4} g^2
   g'^2+\frac{1187 }{216}g'^4+9 g^2
   g_s^2 \nn\\
 & + &  \frac{19}{9} g'^2 g_s^2-108
   g_s^4+\left(\frac{225}{16}g^2+\frac{131 }{16}g'^2+36 g_s^2\right) s
   y_t^2 + 6 \left(-2 s^2 y_t^4-2
   s^3 y_t^2 \lambda +s^2 \lambda   ^2\right)\Bigg{]}.
   \label{betayt}\eea

For the gauge couplings $g_i=\{g',g,g_s\}$ we have
\bea
\beta_{g_i} & = &{1\over(4\pi)^2}g_i^3 b_i+{1\over(4\pi)^4}g_i^3\left[\sum_{j=1}^3 B_{ij}g_j^2-sd_i^t y_t^2\right],
\eea
with
\beq
b=((40+s)/6,-(20-s)/6,-7),\quad
B=\left(
\begin{array}{ccc}
199/18 & 9/2 & 44/3 \\
3/2 & 35/6 & 12 \\
11/6 & 9/2 & -26
\end{array}\right),\quad
d^t=(17/6,3/2,2).
\eeq

Finally,
the anomalous dimension of the Higgs field is
\bea
\gamma &=&  -{1\over(4\pi)^2}  \left[\frac{9 g^2}{4}+\frac{3 g'^2}{4}-3
   y_t^2\right] \nn\\
   &-& {1\over(4\pi)^4} \left[\frac{271
   }{32}g^4-\frac{9}{16} g^2 g'^2-\frac{431
   }{96}s g'^4-\frac{5}{2} \left(\frac{9}{4}g^2+\frac{17
   }{12}g'^2+8 g_s^2\right)
   y_t^2+\frac{27}{4} s y_t^4-6
   s^3 \lambda ^2\right].\,\,\,
\eea


\section{Remarks on Running Inflation as an EFT}
\label{NewApp}

The Lagrangian analyzed in this Letter is not renormalizable in the conventional sense, nor is it ``technically natural" from the point of view of effective field theory. In this appendix we 
remark on the validity of such a theory at high energies (for related discussions see \cite{Burgess,Espinosa}) 
and elaborate on the spirit of our calculations.

The novelty of running inflation is to introduce the non-minimal coupling $\xi\phi^2\mathcal{R}$ into the low energy Lagrangian, which is allowed by all known symmetries of the Standard Model and gravity. This term is dimension 4 in the same sense that the kinetic term $g^{\mu\nu}\partial_\mu\phi\partial_\nu\phi$ is also.
However, if we expand around flat space $g_{\mu\nu}=\eta_{\mu\nu}+h_{\mu\nu}/\mpl$ then the new term is dimension 5 at leading order, plus an infinite tower of corrections
\begin{eqnarray*}
\xi\phi^2\mathcal{R}\sim\xi\phi^2 \Box h/\mpl+\ldots,
\end{eqnarray*}
which is connected to the non-renormalizability of gravity in 4 dimensions. This suggests that non-minimally coupled theories becomes strongly interacting at scales $\Lambda \sim \mpl/\xi$. This can be compared to minimally coupled theories with $\Lambda \sim \mpl$.

Without any protecting symmetry, we cannot forbid infinite towers of corrections to the dimension 4 effective Lagrangian $\mathcal{L}_4$, including those of the form
\begin{equation*}
\mathcal{L} = \mathcal{L}_4 + \lambda\phi^4\sum_{n>0}a_n\left(\frac{\phi}{\Lambda}\right)^n
+\xi\phi^2\mathcal{R}\sum_{n>0}b_n\left(\frac{\phi}{\Lambda}\right)^n+\ldots,
\end{equation*}
which applies to non-minimal models (with $\Lambda\sim\mpl/\xi$) and minimal models (with $\Lambda\sim\mpl$). The values of the higher order Wilson coefficients $a_n,b_n$ cannot be determined without knowledge of the behavior of gravity at energy scales above $\Lambda$, since these terms arise from graviton exchange. If we take a naive estimate $a_n, b_n=\mathcal{O}(1)$, then the required flatness of the inflationary potential is jeopardized. This applies both to running inflation and to many minimal inflation models, such as $m^2\phi^2$ chaotic inflation, since in both cases $\phi>\Lambda$ during inflation. As there is no increased symmetry in the limit $a_n,b_n\to 0$,
such theories are not ``technically natural".

On the other hand, we currently have no evidence for $a_n,b_n=\mathcal{O}(1)$, as these terms arise from graviton exchange, whose effects are yet to be seen in any experiment. There does exist a logical possibility that graviton exchange at high scales is softer than naive estimates suggest (Ref.~\cite{Horava} may be an example), rendering $a_n,b_n$ small, preserving unitarity, and leaving our calculated potential $V(\phi)$ essentially unaltered. It is in this spirit of including only the known Standard Model loops, and not those of  unknown graviton loops, that we have obtained our results and predictions -- which are highly falsifiable.


\end{document}